\documentclass[a4paper,noshowpacs,showkeys,preprint,superscriptaddress,prb,nobibnotes]{revtex4}
\usepackage{graphicx}
\usepackage{amssymb}
\usepackage{amsmath}
\usepackage{epsfig}

\usepackage[usenames]{color}

\usepackage[top=1.5in, bottom=1.25in, left=1in, right=1in]{geometry}

\begin{document}

\title{Comment on "Ab initio calculations of the lattice parameter and elastic stiffness coefficients of bcc Fe with solutes" Comp. Mat. Sci. v.126 pp.503-513 (2017)}

\author{D.~Psiachos}
\email{dpsiachos@gmail.com}

\begin{abstract}
In a recent paper, the authors propose to separately calculate the volumetric and chemical contributions to the elastic stiffness coefficients of systems containing solutes, as
it is ``computationally more efficient". We show that this is not the case and further that their methodology and hence their results are incorrect. There is no short cut
for performing the desired calculations, if done rigorously, as we show in our 2012 work~\cite{2012}.
\end{abstract}

\maketitle
In a recent paper~\cite{Fellinger}, the authors propose to separately calculate the
volumetric and chemical contributions to the elastic stiffness coefficients of systems containing solutes, as
it is ``computationally more efficient" (\textit{c.f.} the abstract). I would like to comment on Fellinger et al's statement on p.504,
which is crucial for understanding the aims of the paper and what it actually accomplishes:
\textit{``Psiachos et al. [7] calculate the total and volumetric changes in the elastic stiffness coefficients of bcc Fe
due to H solutes, and estimate the chemical contribution as the difference between the total and volumetric
changes. We extend this approach by separately calculating the volumetric and chemical contributions to changes in the
elastic stiffness coefficients. We show that the sum of these two contributions agrees with direct calculations
of changes in the elastic stiffness coefficients that encompass both effects..."}

While it is true that the ``chemical" contribution to the elastic parameters is not explicitly calculated
but rather deduced in the 2011 Acta Mat. paper~\cite{ActaMat} referenced in the above statement
([7] in Fellinger et al's reference list), such a calculation is meaningless as the sum of the two interactions - volumetric
and chemical - does not formally lead to the total, ``direct" (Fellinger et al's terminology) effect as I will show. A small difference
in our approaches is that Fellinger et al work with stress-strain coefficients while we use energy-strain coefficients (below, 
as in Refs.~\onlinecite{ActaMat}-\onlinecite{2012}). However, this is an insignificant detail for demonstrating our claim in this comment, as the two forms of coefficient
are identical for $C_{11}$ ~\cite{ActaMat}.

What my 2012 paper~\cite{2012} has in common with the Fellinger \textit{et al} paper is the reliance on elastic theory and series
expansions for treating defects, similar to but more thorough than the 2011 Acta Mat. paper~\cite{ActaMat}. In the 2012 paper~\cite{2012}, the ``chemical" contribution
can come about by combining Eqs. 5 and 7 with $Y=x$ (or $c_s$ in the notation of Fellinger et al) and $V^{ref}=V^0$. $dC^{tot}/dx$ would correspond
to the ``direct" term in their notation (\textit{viz.} Fellinger et al's Eq. 11).
 If this is done and $dC^{tot}/dx$ (about the undefected volume $V^0$) is calculated one obtains the ``chemical" part, which is the same as
Fellinger et al's Eq.10, plus a volumetric-derivative term of the defected system, which is not taken about $V^0$ but about the defected-system
equilibrium volume, which causes it to \underline{not be constant}.

Briefly, using Eqs. 5 and 7 of my 2012 paper~\cite{2012},
\begin{eqnarray*}
\left.\frac{dC^{tot}}{dx}\right|_{V^0}&=&\left(C^x(V^x)-C^0(V^0)\right)/x\\
&=&\frac{C^x(V^0)-C^0(V^0)}{x}+\frac{1}{V^x}\frac{dV}{dx}\left.\frac{\partial C^{x}}{\partial \eta}\right|_{V^x}.
\end{eqnarray*}

The first term, the chemical part, is the same as in Fellinger et al (Eq.10). However, in the second term, the volumetric derivative about $\eta$ pertains
to the defected system and is taken about that system's equilibrium volume, not $V^0,$ the undefected volume. This should be contrasted
with Fellinger et al's Eq. 5 which is the volumetric derivative of the
undefected system about the undefected volume with lattice parameter $a_0$.

In the above, the notation is defined in~\onlinecite{2012} but briefly, $x$ corresponds to a system with 
solute concentration $x,$ the superscript $0$ pertains to the undefected system, 
$C$ are the elastic constants, $V$ is the volume, and $\eta$ is the volumetric strain. The superscript `\textit{tot}' corresponds to the ``direct"
term of Fellinger et al. (their Eq.11). 

Therefore, the sum of the``volumetric" and ``chemical"
components as defined by Fellinger \textit{et al} do not exactly equal the result of the ``direct" calculation $dC^{tot}/dx$ in Ref.~\onlinecite{ActaMat} or \onlinecite{2012}, a
quantity which is equivalent to Eq. 11 in theirs.
In summary, while one can formally split up the ``direct" term into a ``chemical" and ``volumetric" contribution and calculating
only these, doing so confers no computational advantage over calculating the direct term given 
that volumetric derivatives of defected systems need to be calculated as part of the volumetric contribution. In summary, the methodology
and hence the results presented in this paper are incorrect. Only the rigorous derivation in Ref.~\cite{2012} can lead
to correct outcomes, and there is hence no short-cut to performing these calculations.

\end{document}